\documentclass[preprint,amsmath,amssymb,aip,apl,floatfix]{revtex4-1}
\usepackage{amsmath}
\usepackage{graphicx}
\usepackage{dcolumn}
\usepackage{bm}
\usepackage[dvipsnames]{xcolor}
\usepackage{float}

\begin{document}

\title{A comparison of two different mechanisms for deterministic spin orbit torque magnetization switching}

\author{F. Garcia-Sanchez}
\email{f.garcia@inrim.it}
\altaffiliation{Current address: Departamento de Física Aplicada, Universidad de Salamanca, Pza de la Merced s/n, 37008 Salamanca, Spain}
\affiliation{Istituto Nazionale di Ricerca Metrologica (INRIM), Strada delle Cacce 91, 10135 Torino, Italy}
\author{G. Soares}
\affiliation{Istituto Nazionale di Ricerca Metrologica (INRIM), Strada delle Cacce 91, 10135 Torino, Italy}
\affiliation{Department of Applied Science and Technology (DISAT), Politecnico di Torino, 10129 Torino, Italy}
\author{M. Pasquale}
\affiliation{Istituto Nazionale di Ricerca Metrologica (INRIM), Strada delle Cacce 91, 10135 Torino, Italy}

\date{\today}

\begin{abstract}
In this article we analyze by modeling two possible mechanisms for magnetization switching using spin orbit torques, which have been reported to cause field-free deterministic switching in experiments. Here we compare the field-free magnetization switching due to a tilt of the anisotropy direction against the use of an antiferromagnetic bias field. Simple results obtained analytically show that a bias field not only causes the magnetization reversal but also reduces the corresponding energy barrier. The critical current required for magnetization switching is analyzed on the basis of a macrospin model. It is shown that although the field-free deterministic switching caused by a tilt of the anisotropy is more robust than the bias field in the development of memory elements, a compromise between requirements has to be adopted when selecting the parameters for specific applications.
\end{abstract}

\maketitle
\section{Introduction}

Magnetic recording devices have always been of paramount importance to society, as they supplied and still supply the demand for higher and higher data storage capacity. Novel magnetic phenomena like spin transfer torque (STT) have also been successfully used to produce memory devices such as the non-volatile magnetic random access memory (MRAM)\cite{apalkov2016,kent2015}.

Spin orbit torques \cite{miron11,liu12,garello2014,reviewsot,manchon2019} (SOTs) are an effective way to switch the magnetization of perpendicularly magnetized thin nanoelements by means of electric currents. Imposing an electric current on a heavy metal layer (HM), a spin current is generated by the spin Hall effect and its moments can be transferred to an adjacent ferromagnetic (FM) layer, causing a torque on its magnetization. Alternatively, the electric current generates an accumulation of polarized spin density at the interface, causing a torque on the adjacent FM layer by exchange interaction\cite{sinova2015}. These phenomena are of great interest for spintronic devices, where they can potentially substitute the present writing systems - especially for racetrack memories and magnetic logic devices. In such applications, there are size scalability advantages by using perpendicular anisotropy materials (PMA), where the minimum current needed to switch the elements does not depend on the shape anisotropy, a non-thermal contribution to stability. However, to achieve deterministic magnetic switching in PMAs, one needs to perturb the energy landscape\cite{mangin2006} of the system. The simplest way to perturb the energy and obtain SOT switching uses an external bias field. Unfortunately, although deterministic switching has been successfully attained in \cite{miron11,liu12_2,liu12,avci12,avci14,jiang2019}, applying an external in-plane (IP) field undermines the technological advantages of utilizing SOT for practical purposes.

\begin{figure*}[htb]
\includegraphics[width=\textwidth]{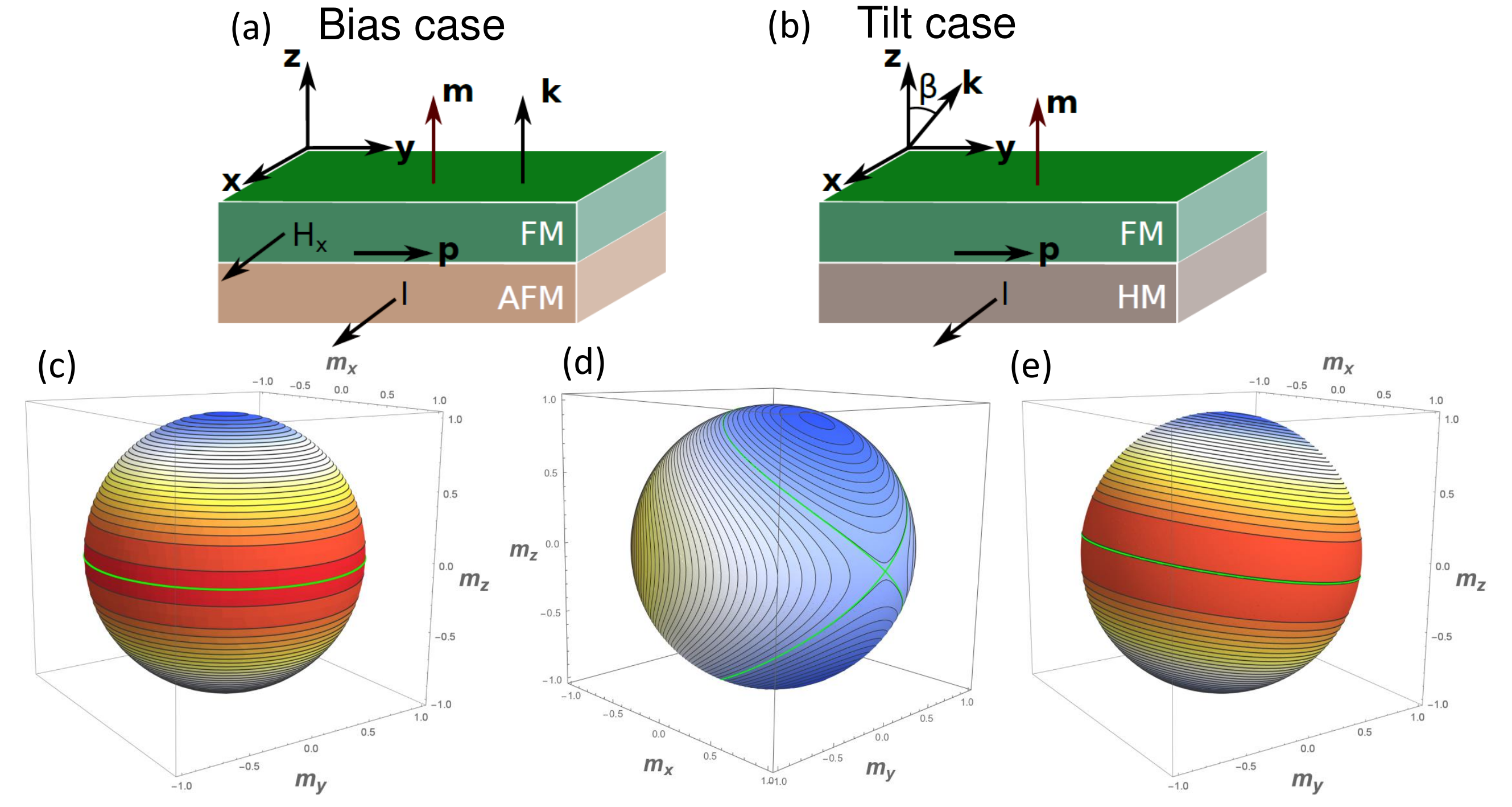}
\caption{(a) Vector diagram of the bias case and (b) tilt case. The anisotropy tilt angle is in the yz plane for the tilt case. Energy landscape for (c) Perpendicular anisotropy, (d) Bias case and (e) Tilt case. In the graphs, the blue color is the minimum value and red is the maximum value. The magnetization may switch if the applied current is sufficient to traverse the separatrix line plotted in green. Energy-conserving trajectories are shown in black.}
\label{Fig:landscape}
\end{figure*} 

In the context of practical use for SOT-MRAM three terminal devices \cite{mangin2006,jabeur2013,cubuckcu2014,cubucku2018,manchon2019}, where the magnetic-tunnel junction readout block is ferromagnetically uncoupled, a minimum stacking of layers is desired. A first solution compatible with SOT switching is to generate an IP bias field adding an extra magnetic layer: either an uncoupled FM layer on the stack or an HM antiferromagnet layer such as PtMn\cite{fukami16,kurenkov} or IrMn\cite{Oh}, two materials which can simultaneously provide the spin current and the necessary bias field. A second solution to obtain SOT switching, is to modify the energy landscape by breaking the spatial symmetry of the PMA energy barrier, which introduces an angular tilt in the out-of-plane (OOP) anisotropy axis. Torrejon et al.\cite{torrejon15} and You et al.\cite{you2015} engineered the anisotropy tilt using film-thickness gradients on either the FM or the MgO oxide layer, and obtained a field-free deterministic switching on HM/FM/MgO trilayers. 
 
It is worth mentioning that other field free switching methods have been recently proposed to achieve the system symmetry-breaking. Some of them are: extra in-plane shape anisotropy due to a elongated FM shape \cite{wang2019} and extra input current terminal on the \textbf{x} direction \cite{deorio2019,wang2019,sverdlov2019}; extra field-like torque caused by a thickness gradient of the oxide layer \cite{yu2014,cui2019}; RKKY coupling with a second FM layer \cite{tremsina2019,Lau2016}; laterally attached in-plane anisotropy FM structures \cite{long2015}, switching by induction of domain nucleation \cite{murray2019} ; the creation of a T type structure\cite{kong2019}; composite free-layer composed by antiferromagnetically coupled layers \cite{li2019} and Co/Pt/Co trilayers with orthogonal easy axis between Co layers\cite{Lazarski2019}. Nonetheless, these methods cannot be implemented easily or without the introduction of additional layers in the three-terminal architecture device. As a matter of technological applicability, the in-plane bias field or anisotropy tilt methods remain of great interest and potential.
 
Until now, it is still unclear if an in-plane bias field or an anisotropy tilt would only help the SOT switching or would also interfere with the energy barrier height and critical current values: in the scenario of a reduction of the energy barrier, the possible miniaturization of the devices would be jeopardized, because the memory elements would become thermally unstable.
Since the relationship between the energy barrier and the critical current is extremely important for SOT based data storage applications, it needs to be investigated thoroughly.

To this end, in this work we focus on the writing performance of both methods described above to achieve field-free SOT magnetization switching. Namely, a fixed bias field such as the AFM/FM bilayers (bias case) and an anisotropy tilt caused by a structural asymmetry (tilt case). The writing performance is analyzed with respect to two requirements of magnetic recording: a large energy barrier for thermal stability and a small critical current for efficient switching. As such, the results may be useful independently of the chosen reading method for specific three-terminal applications.
We start investigating analytically the effects of a bias field induced by an AFM/FM layer or an anisotropy tilt using an HM/FM system on the energy landscape. We then proceed to solve numerically the Landau-Lifshitz-Gilbert equation on a macrospin approximation either at several different in-plane field values or with different anisotropy tilt angles, interpreting the results in the light of the effective bilayer energy landscape. Finally, we investigate the effect of the current pulse duration, equivalent to the speed of operation of an hypothetical SOT switching device.

\section{Energy barrier modeling}

We start the discussion by presenting a simple model for the two mechanisms (see Fig.\ref{Fig:landscape}(a) and (b)).
In both cases, the systems posses perpendicular anisotropy. The energy density of a perpendicular anisotropy system is 
$E/V=-K_{u}m_z^2$ where $K_{u}$ is the effective uniaxial anisotropy constant, $m_z$ is the third component of the normalized magnetization $\vec{m}$ and $V$ is the volume of the ferromagnetic layer. Here, the anisotropy constant is the resultant of the out-of-plane anisotropy $K_{oop}$ and the shape anisotropy $K_{u} = K_{oop} - \mu_0M_s^2/2$, where $\mu_0$ the vacuum permeability and $M_s$ the saturation magnetization. We are considering PMA systems and, thus, $K_{u}$ is strictly positive. The energy landscape is symmetric as it is shown in Fig.\ref{Fig:landscape}(c). The saddle point corresponds to the equator and the minima to the poles of the sphere.

We consider the bias case a ferromagnet in contact with an antiferromagnet as depicted in Fig.\ref{Fig:landscape}(a). In a simple approximation we can model it with a unidirectional field, representing the strength of the exchange field bias. The energy density of the system becomes:
\begin{equation}
E/V=-K_{u}m_z^2-\mu_{0}M_{s}m_xH_{X}
\label{Eq:energyafm},
\end{equation}
where $H_{X}$ is the exchange bias field. The energy landscape corresponding to such energy density is shown in Fig.\ref{Fig:landscape}(d).
 
In the other mechanism, which we label as the tilt anisotropy case, we model the energy by adding a tilt angle on the anisotropy axis (see diagram in Fig.\ref{Fig:landscape}(b)). Under this assumption the energy density becomes:
\begin{equation}
E/V=-K_{u}(\vec{m} \cdot \vec{k})^2
\label{Eq:energytilt},
\end{equation} 
where $\vec{k}$ is the anisotropy axis and has the form $\vec{k}=(0,sin\beta ,cos \beta)$ where $\beta$ is the tilt angle. We plot the landscape of such case in Fig.\ref{Fig:landscape}(e) (right). The equilibrium state is aligned to the anisotropy axis and then $\vec{m}_{eq}=(0,\pm sin\beta ,\pm cos \beta)$.

\subsection{Equilibrium positions} In the bias case, when the bias field $H_X$ is smaller than the anisotropy field $H_{k}=2K_{u}/\mu_{0}M_{s}$, the system has two equilibrium positions with positive and negative projection on the z axis (top and bottom). Their values in polar coordinates are:
\begin{equation}
\theta_{eq}=\arcsin(\frac{H_{X}}{H_{k}})
\label{Eq:angle}.
\end{equation}
 These equilibrium positions correspond to the linear part of the hard axis IP hysteresis loop with perpendicular anisotropy, with $m_{x}=H_{X}/H_{k}$. There are two valleys corresponding to such minima as shown Fig.\ref{Fig:landscape}(d). There is also a high energy region, which contains the global energy maximum. This region corresponds to the white and yellowish regions of Fig.\ref{Fig:landscape}(d). Both the minima valleys and the high energy regions are partitioned by the separatrix (green line) as can be seen in Fig.\ref{Fig:landscape}(d). In the anisotropy tilt case the separatrix is rotated by the tilt angle and there are no other equilibrium regions apart from the two minima (see Fig.\ref{Fig:landscape}(e)). The existence of the high energy region in the bias case will change the behavior of that system. In the anisotropy tilt case, once the current is switched off the magnetization will evolve to the minimum corresponding to the nearest valley and will relax to the corresponding energy minimum. In the bias case, if at the end of the current pulse the magnetization is in a minimum valley, the magnetization will spontaneously evolve to the corresponding minimum, due to energy relaxation. However, if at the end of the current pulse, the magnetization is in the high energy region, the final state of the magnetization will depend on the exact point of the high energy region where it was at the time of current removal. This will result in a bias case switching diagram more dependent on the values of the parameters (current value or pulse duration) as will be discussed in the following sections.

\subsection{Energy barriers of the two systems} 
The simplest thermal switching mechanism corresponds to coherent rotation: the rotation of the magnetization as a whole, which is similar to the one described by a macrospin approach. To obtain the energy barriers of the system one needs to determine the position of the saddle points. In the bias case, the saddle points correspond to the polar coordinates $\phi=0$ and $\theta=\pi/2$, associated to the crossing of the separatrix line (see Fig.\ref{Fig:landscape}(d)). The energy barrier $E_B$ for coherent rotation then becomes: 
\begin{equation}
E_{B}=\frac{\mu_{0}M_{s}V(H_{k}-H_{X})^{2}}{2H_{k}}=K_uV\left(1-\frac{H_{X}}{H_{k}}\right)^{2}.
\label{Eq:barriercoherent}
\end{equation}

\begin{figure}[hb]
\includegraphics[width=0.5\textwidth]{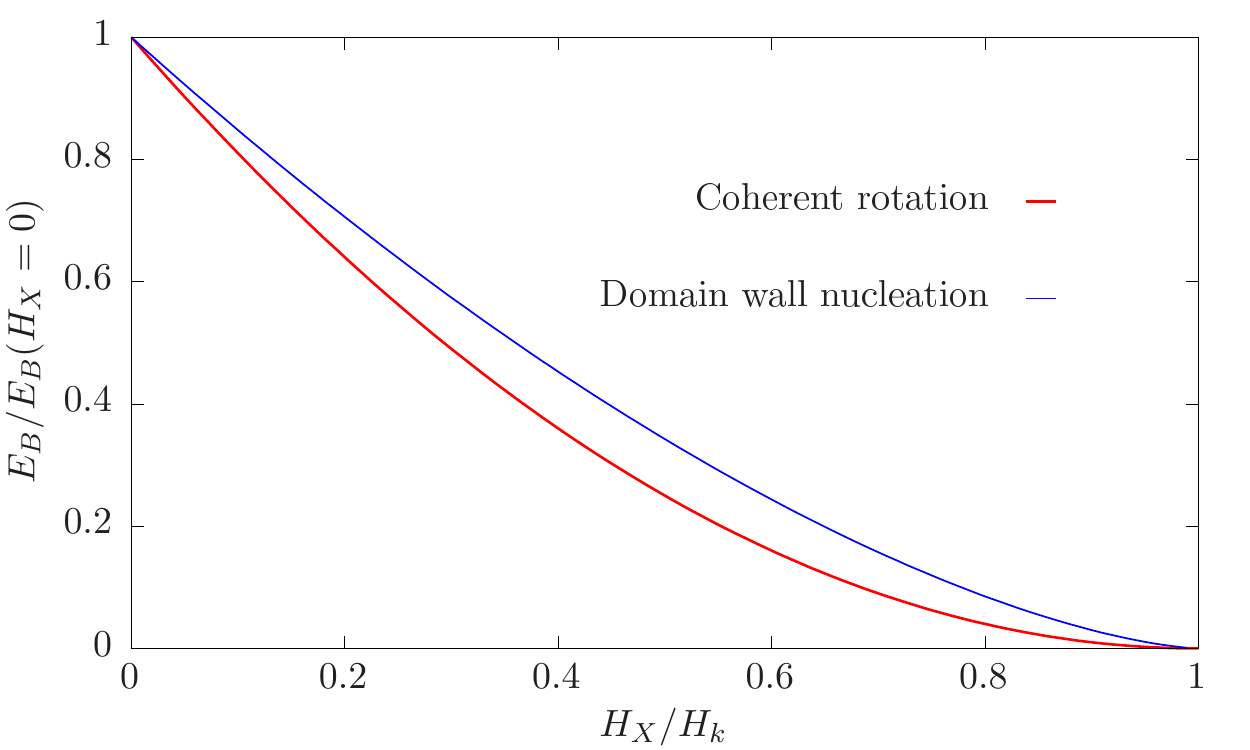}
\caption{Energy barrier for coherent rotation and domain wall nucleation as
a function of the exchange-bias field.}
\label{Fig:ebarrier}
\end{figure}

\begin{figure*}
\includegraphics[width=\textwidth]{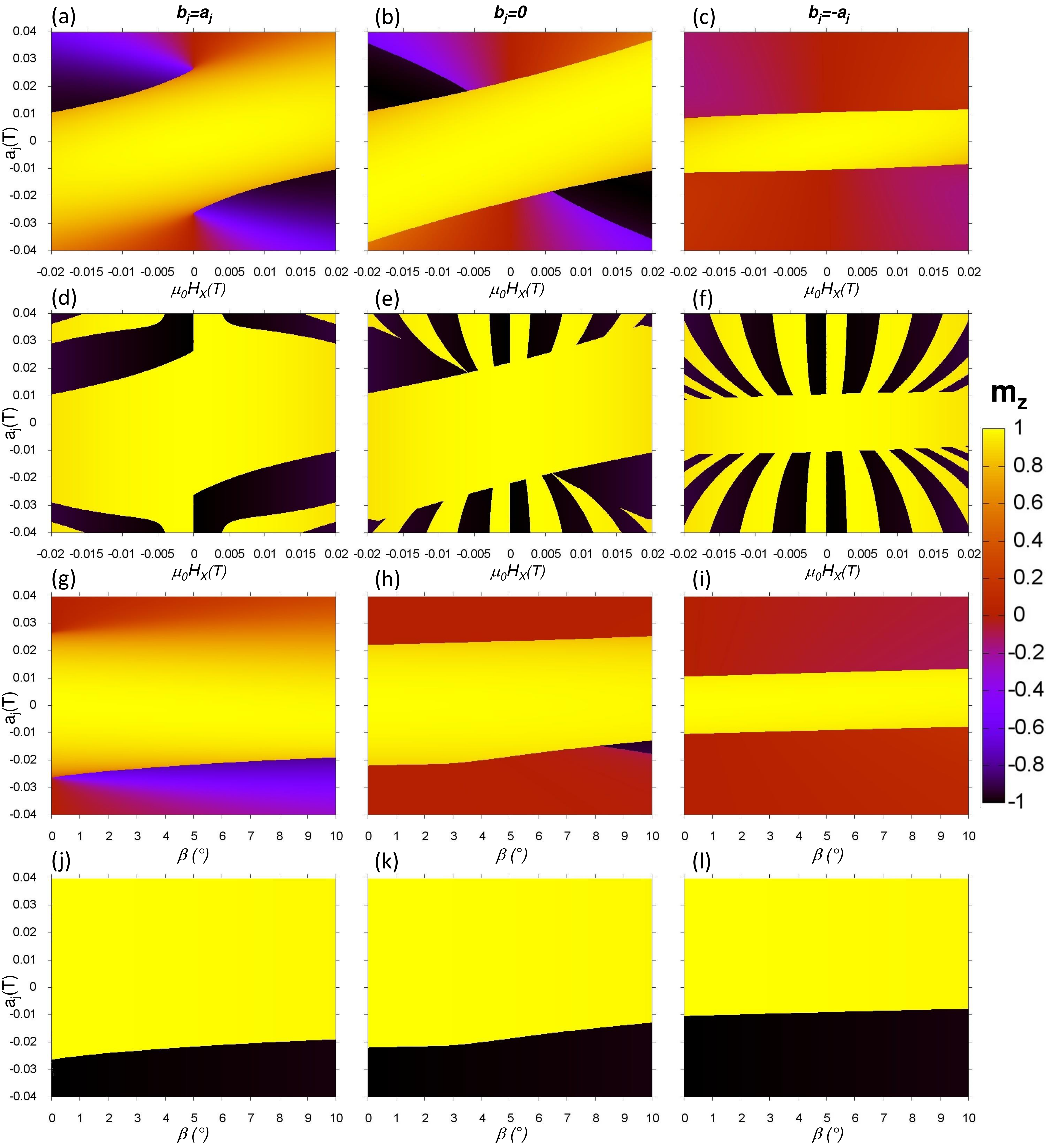}
\caption{OOP magnetization map ($m_z$ color coded) with different $b_j$ values $b_j=a_j$; $b_j=0$; $b_j=-a_j$ as a function of the intensity $a_j$ of a current pulse lasting $t_p$: from a) to c) $m_z$ is the bias case as a function of the bias in plane field ($H_{X}$) at the end of current pulse time $t_p=100 ns$ , d) to f) the relaxed $m_z$ for the bias case after 100 ns ($t=2*tp$) upon removal of the current pulse; g) to i) Anisotropy tilt case as a function of the anisotropy tilt angle $\beta$ at the end of current pulse time $t_p=100 ns$, j) to l) the relaxed $m_z$ for the tilt case after 100 ns ($t=2*tp$) upon removal of the current pulse;. Column-wise we observe the effect of the field-like torque $b_j$ in the systems.
}
\label{Fig:combined_sw}
\end{figure*}

The energy barrier for coherent rotation, when there is no bias field, yields the typical value $E_{B}=K_uV$. Using an effective anisotropy constant $K_{u}=2.65\times 10^4 J/m^3$ and an element of volume 100x100x1 $nm^3$, the thermal stability factor has the value $E_{B}/k_BT_{Room}=64$. This value is close to the standard value $E_{B}/k_BT_{Room}=60$ in magnetic recording\cite{weller99} for thermal stability of ten years. In the anisotropy tilt case, the energy barrier height for coherent rotation (as well as the stability factor) is maintained because the tilt corresponds to a rigid rotation of the energy landscape without variations of its value. From the comparison of the coherent rotation value in zero field with Eq.\ref{Eq:barriercoherent}, one can conclude that the introduction of an antiferromagnetic layer, which produces a bias field orthogonal to the anisotropy axis, tends to reduce the energy barrier by a factor of $(1-H_{X}/H_{k})^2$. 

\subsubsection{Energy barriers for nucleation of a domain wall}

The thermal switching process which is most likely to occur is the process with the minimum energy barrier. As we increase the size of the element, volume grows, and other switching processes with a smaller energy barrier become available such as the thermal switching by nucleation and propagation of a domain wall. This process can not be described by the macrospin model, but its energy barrier can be calculated by adding the exchange energy $A[(\nabla m_x)^2+(\nabla m_y)^2+(\nabla m_z)^2]$ term to the system energy, where $A$ is the exchange constant.
In the case of the energy barrier for the nucleation of a domain wall, $E_B$ corresponds to the domain wall energy, which in absence of a bias field is $E_{B}=4S\sqrt{AK_{u}}$, where $S$ is the smallest cross section of the device. The saddle point configuration for that energy barrier corresponds to a $180^\circ$ wall. For an element of cross section $S=100 nm^2$, an effective anisotropy constant $K_{u}=2.65\times 10^4 J/m^3$ and exchange constant $A=1.5\times 10^-11 J/m$, the thermal stability factor becomes $E_{B}/k_BT_{Room}=60.88$. The value of the energy barrier in the anisotropy tilt case is preserved because the saddle point configuration is also a $180^\circ$ wall. In the bias case, when a bias field is present, the domains are tilted and the energy of the domain wall is reduced due to the smaller rotation of the magnetization inside the wall. The saddle point configuration then contains a domain wall of $\pi-2\theta_{eq}$ as defined in Eq. \ref{Eq:angle}. The energy barrier for a domain wall in the bias case is given by:

\begin{equation}
\begin{split}
E_{B}=4S\sqrt{AK_{u}}\left(\sqrt{1-\left(\frac{H_{X}}{H_{k}}\right)^{2}} \right.\\ \left.+\frac{2H_{X}}{H_{k}}\arctan(\frac{H_{X}-H_{k}}{\sqrt{H_{k}^{2}-H_{X}^{2}}}) \right)
\end{split}
\label{Eq:barrierdw}
\end{equation}
which is equivalent to the formula already found in literature \cite{hubert98}. The energy barrier for coherent rotation and domain wall nucleation is shown in Fig.\ref{Fig:ebarrier} as a function of the exchange bias field $H_{X}$. In both cases the energy barrier is reduced with respect to the tilt case (zero bias field) case. The energy barrier has to be calculated more precisely in the general case, where inhomogeneities of the magnetization can play a role. 

For the bias case, the domain wall width of the saddle configuration is field dependent and has the value
\begin{equation}
\Delta=\frac{\Delta_{0}(H_{X}+H_{k})}{\sqrt{H_{k}^{2}-H_{X}^{2}}}
\label{Eq:dwwidth}
\end{equation}
where is $\Delta_{0}=\sqrt{(A/K_u)}$.
For such calculations the criterion of the derivative at the middle of the wall has been used.
The thin films comprised of multilayers of HM/FM can also posses a chiral exchange of interfacial origin known as Dzyaloshinskii-Moriya interaction (DMI)\cite{fert90}. In the presence of interfacial DMI the energy barrier corresponding to a DW nucleation will be reduced by a factor $-\pi DS$ \cite{Pizzini14} where $D$ is the DMI constant. This is in agreement with some numerical calculations showing that DMI actually reduces the energy barrier for switching in magnetic nanoelements \cite{cubuku16,gastaldo2019}.

In summary, the introduction of a bias field or DMI reduces the energy barrier, while the anisotropy tilt preserves it, independently of the type of thermal switching mechanism. However, the thermal mechanism does not determine the dynamical switching and has to be analyzed separately with numerical modeling. For this reason we explore the dynamics under applied current in the next sections.

\section{Critical currents}
 
Following the analytical discussion in the previous section we now investigate the effect of a current pulse on the SOT device using numerical calculations of the dynamics of magnetization as a function of the applied current induced damping like torque $a_j$ and pulse time duration $t_p$. We use the same dynamics and torque equations for both the bias case and anisotropy tilt case, changing only the variable of interest: the bias field $H_{X}$ for the bias case and the tilt angle $\beta$ for the anisotropy tilt case (see Fig.\ref{Fig:landscape}(a) and (b)). We focus on identifying the minimum necessary value of the dc current generated damping like torque $a_j$ to obtain SOT switching, i.e. the critical current. We then interpret the results in light of the energy landscape obtained in the previous section\footnote{We have chosen to use the current torque in field units, since the macrospin problem can be normalized to the anisotropy field values and the actual value depends on the related material conditions.}. 

The dynamics of the magnetization was calculated using the Landau-Lifshitz-Gilbert equation with the additional terms corresponding to the field-like and damping-like spin orbit torques
\begin{equation}
\frac{d\hat{m}}{dt}=-\gamma \hat{m} \times (\vec{H}_\text{eff}+a_j(\hat{m}\times\hat{p})+b_j\times\hat{p})+\alpha \hat{m} \times\frac{d\hat{m}}{dt}
\label{Eq:llg}
\end{equation}
where $\gamma$ is the gyromagnetic ratio and $\alpha$ is the Gilbert damping parameter.
$\vec{H}_\text{eff}=-(1/\mu_0 M_s)\partial (E /V)/\partial\vec{m} $. We also use here the macrospin approach, where the magnetization is considered to be homogeneous in the magnetic layer. This approach has been shown to be valid as long as no strong DMI\cite{martinez2015} is present or large applied currents are used\cite{finocchio2013}. For such situations, the switching is mediated by inhomogeneities like domain walls\cite{martinez2015} and full micromagnetic simulations are required. The effective field contains the anisotropy field and in the bias case, the bias field is introduced inside $\vec{H}_\text{eff}$ as an additional external field $H_{X}\hat{x}$. We consider the polarization direction of the spin Hall current to be $\hat{p}=(0,1,0)$. 
The ratio between field-like torque induced by the $b_j$ current and damping-like torque connected to $a_j$ is $b_j/a_j$ and depends on the material; since in AFM/FM bilayers (bias case) it has been reported to be affected by the order of the stack \cite{ou16}, while in HM/FM bilayers (tilt case) it depends on interface effects on the ferromagnet, we have considered for simplicity three possible scenarios for the field like torque $b_j$: $b_j=a_j$; $b_j=0$ and $b_j=-a_j$. The current pulse time and subsequent relaxation is governed by the step-like Heaviside function, since typical rise and fall times of pulse generators are in the range from about $20$ to $100$ ps and no capacitive or reactive phenomena are considered. The anisotropy constant corresponds to $\mu_0 H_{k}=0.053$ $T$ as in Ref.\cite{torrejon15}. The damping value, which can vary due to the different interfaces of the bias and anisotropy tilt cases, was anyhow fixed to be $ \alpha= 0.05$ to be able to compare the various results.

In order to determine the critical current values for the SOT switching in the tilt and bias cases we start with the magnetization in the equilibrium state corresponding to the positive Z direction, then we apply a single $t_p=100$ ns current pulse $a_j$ and the results are shown in Figs. \ref{Fig:combined_sw}(a-c) bias case and \ref{Fig:combined_sw}(g-i) tilt case. The stability of both cases are also analyzed by letting the magnetization relax after the 100 ns pulse in Fig. \ref{Fig:combined_sw}(d-f) for the bias case and \ref{Fig:combined_sw}(j-l) for the anisotropy tilt case. 

\subsection{Anisotropy tilt case} For the anisotropy tilt case, we vary the amplitude of the $t_p= 100$ ns current pulse $a_j$ and the angle $\beta$ while fixing for simplicity the $b_j$ values as described in the previous paragraph. In Fig. \ref{Fig:combined_sw}(g-i) we show the magnetization state right after the end of the pulse duration $t_p$. Here, since the energy landscape is not distorted but only rigidly rotated by an angle $\beta$ (see Fig.\ref{Fig:landscape} (e)), the switching is rather robust. 

\begin{figure}[ht]
\includegraphics[width=1\textwidth]{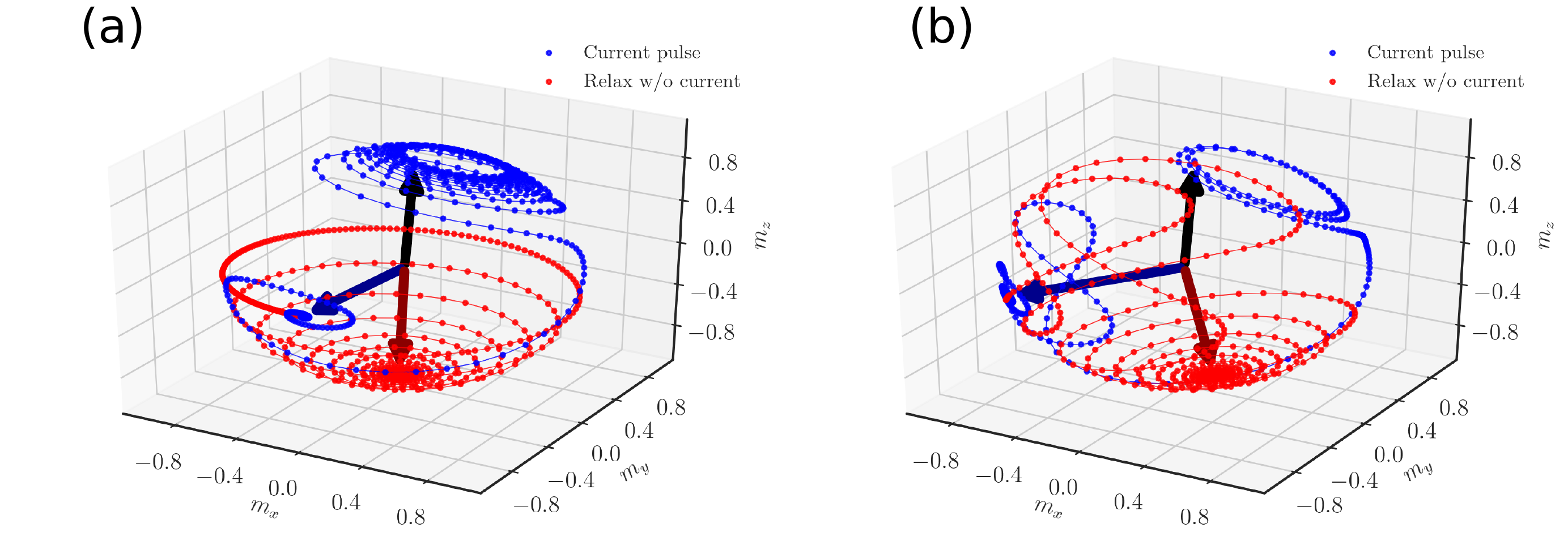}
\caption{Simulation of the magnetization trajectory for field free switching with $a_j=-10$ mT and $b_j=-a_j$. Black arrow indicates the initial state, blue lines and dots dynamics with the current pulse and red lines and dots the relaxation upon removing of the current. a) Tilt case, $\beta=3.5^o$. b)Bias case, $H_{X}=-10$ mT. }
\label{Fig:trajectory}
\end{figure}

The damping like torque caused by the applied current ($a_j$) pulls the magnetization to the plane of the sample ($m_z=0$), and may be assisted by the field like torque $b_j$ or the anisotropy tilt $\beta$ depending on its respective directions (see blue trajectory on Fig.\ref{Fig:trajectory}(a)) . 

For $b_j=0$ (Fig.\ref{Fig:combined_sw}(h)), about 0.02 T of $|a_j|$ is enough to redirect the magnetization to the plane, and it is assisted by the anisotropy tilt for negative values. The necessary current for the redirection decreases increasing the tilt at negative values, and on the contrary increases for positive values. Meanwhile for $b_j=a_j$ the field like torque also opposes the switching, as it can be seen for the higher currents of Fig.\ref{Fig:combined_sw}(g). On the contrary, the field like torque works in favor of the switching for $b_j=-a_j$ (Fig.\ref{Fig:combined_sw}(i)).

With the magnetization on the plane of the sample, once the current pulse is removed, the system is in a high energy state, and it has to relax to a low energy one. If the adequate conditions are met, the magnetization can then evolve from $m_z \approx 0$ to $m_z=-1$ (red trajectory on Fig.\ref{Fig:trajectory}(a) . The bottom row of Fig\ref{Fig:combined_sw} (j-l) shows the system state after the removal of the current pulse and subsequent relaxation. The switching occurs only for negative values of $a_j$, since in this direction it is bolstered by the rigid energy landscape rotation caused by the tilt, while for negative values no switching happens because for that current polarity, tilt and current torque work in opposite directions. The minimum absolute values of $a_j$ for the switching, in other words the critical current, increases from positive to negative values of $b_j$, suggesting that the field like torque $b_j$ assists the switching when it has a direction antagonistic to the damping like torque Fig\ref{Fig:combined_sw}(l).

\subsection{Bias case}
In the numerical calculations of the bias case, we vary the applied current $a_j$ and the bias field $H_{X}$. In Fig.\ref{Fig:combined_sw}(a) to (c) one sees the final magnetization $m_z$ right after the $t_p$ pulse. 

After the current pulse, the magnetization diagrams resemble to the ones obtained in the tilt case. Opposite signs between the bias field $H_X$ and damping like torque $a_j$ are needed to pull the magnetization in-plane. The field like torque $b_j$ influence, besides $b_j=-a_j$ (Fig. \ref{Fig:combined_sw}(c)), is different though. Generally less current is required to pull the magnetization in the film plane and it can even achieve switching with the help of $|H_X|$, as verified by the black regions of Figs.\ref{Fig:combined_sw}(a) and \ref{Fig:combined_sw}(b). 

However, after the removal of the current pulse, the relaxation is not so straightforward as in the tilt case. Since the energy landscape is distorted by $H_X$, the magnetization has to be also in condition to follow the right path through the lower energy landscape valleys, as evidenced by the lines in Fig.\ref{Fig:landscape}(d) and the red trajectory of Fig\ref{Fig:trajectory}(b). For $b_j=a_j$ and $b_j=0$ (Figs.\ref{Fig:combined_sw} (d) and e)), if the system has already switched it still holds, but not necessarily if the magnetization has to evolve from close to the in-plane condition. Surprisingly, regions that were unlikely to switch do so, as the ones with same signs of $H_X$ and $a_j$. On the other hand, if the magnetization is close to in-plane, the system does not follows the simple continuous behavior, which was verified before on the tilt case, and alternating regions with successful and unsuccessful switching are found.

This is the consequence of a complicated interplay between the spin orbit torques and the current pulse duration, which has been already pointed out by Mangin et al.\cite{mangin2006} and discussed by Miron et al.\cite{miron11}. During the precession period, each of the spin orbit torques ($a_j$ and $b_j$) changes their directions in reference to the magnetization. Thus their actual role in either aiding or obstructing the magnetization switching will depend not only on their intensity and relative directions, but also on the state of the magnetization itself at any given instant. In the simulations just discussed above, large current-pulse-times were used to guarantee a quasi-stationary behavior (net torque equal to zero). The effect of the pulse duration will be further analyzed in section \ref{sec:pduration}.

\begin{figure}
\includegraphics[width=0.5\textwidth]{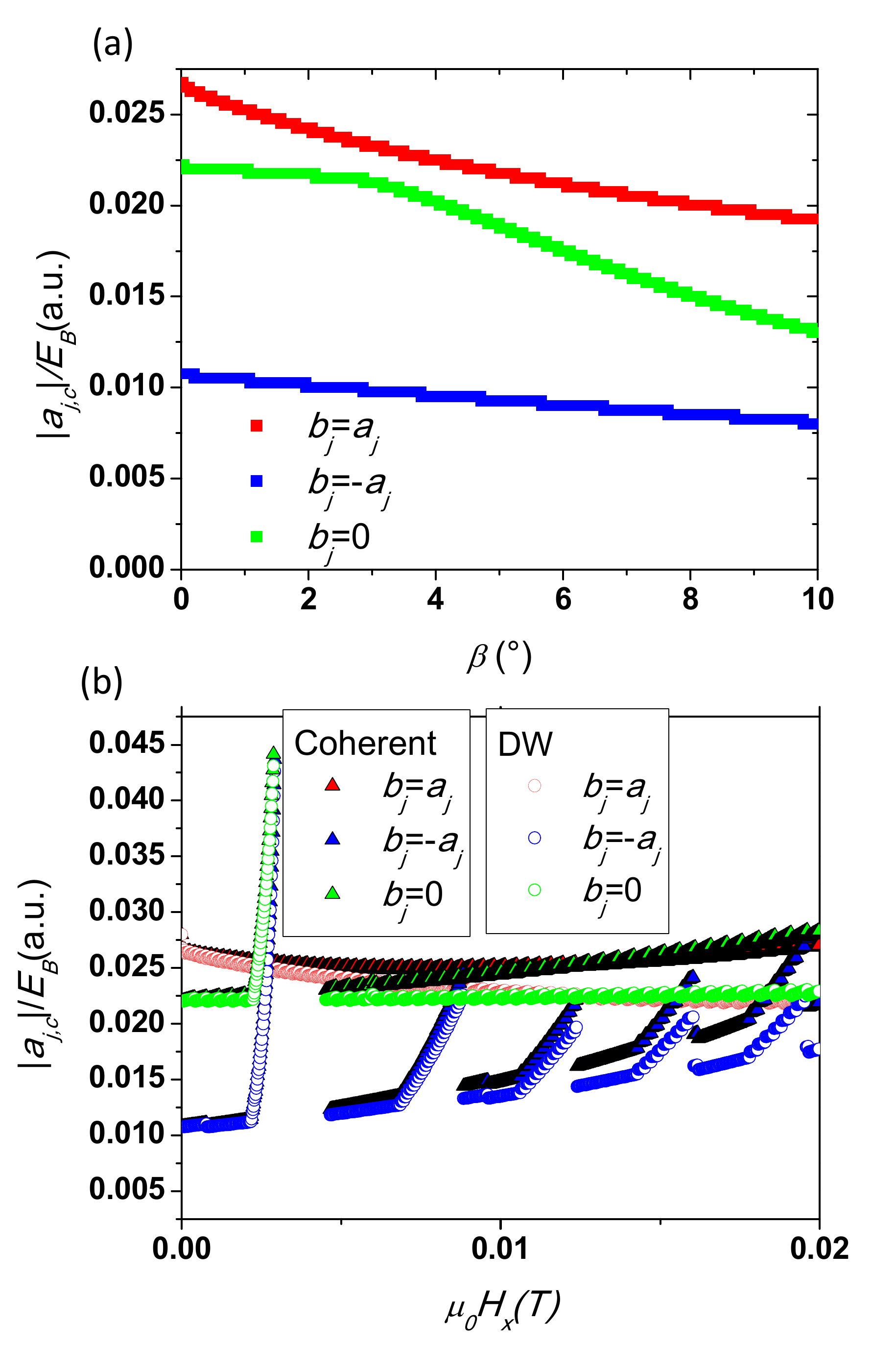}
\caption{Figure of merit $a_{j,c}/E_{B}$ for (a) Tilt case as a function of the tilt angle $\beta$ and (b) Bias case as function of the bias field $H_X$. }
\label{Fig:fmerit}
\end{figure}

\subsection{SOT switching efficiency}\label{efficency} 
Looking at the critical current values $a_{j,c}$ associated to the sign changing of $m_z$ in Fig.\ref{Fig:combined_sw}, it may seem that the bias case is a more energy efficient method to achieve SOT magnetization switching for memory applications. 

Nevertheless, to properly compare the bias and anisotropy tilt switching methods, one needs to normalize the value of the critical current $a_{j,c}$ applied during the $t_p = 100$ ns pulse, used to reach the final relaxed state with respect to the energy barrier $E_B$ encountered in the switching process as defined in Eqs. \ref{Eq:barriercoherent} and \ref{Eq:barrierdw}. To this end in Fig. \ref{Fig:fmerit} we plot the ratio between the module of the critical current and the relevant energy barrier ($|a_{j,c}|/E_B$) as a figure of merit representing the efficiency of both the bias and the tilt cases. 

Comparing Figs.\ref{Fig:fmerit}(a) and (b), in the tilt case we find an evident decrease of the ratio $|a_{j,c}|/E_B$ when the tilt angle $\beta$ increases; on the contrary, an increase of the ratio $|a_{j,c}|/E_B$ is observed when $H_{X}$ increases in the bias case. 
Since in both cases there is a reduction of $a_{j,c}$ with increasing values of the relevant input parameter $H_{X}$ or $\beta$ as shown in \ref{Fig:combined_sw}, we can deduce that the decrease in the energy barrier height $E_B$ with increasing $H_{X}$ in the bias case, must be larger than the reduction of the critical current $|a_{j,c}|$ with $H_{X}$. 

This means that in the bias case we observe an important degrading in the energy barrier. Although this still corresponds to smaller energy requirements on the SOT switching of a possible device, it also entails a lower thermal stability and limits the degree of miniaturization. On the other hand, a larger $|a_{j,c}|$ current needed to switch the magnetization direction in the tilt case would also generate more heat by Joule effect, and this might compromise miniaturization as well. So specific optimization choices are needed in each case for the tailored development of specific SOT applications. 

\section{Switching dependence on pulse duration\label{sec:pduration}}
\begin{figure*}[]
\includegraphics[width=1\textwidth]{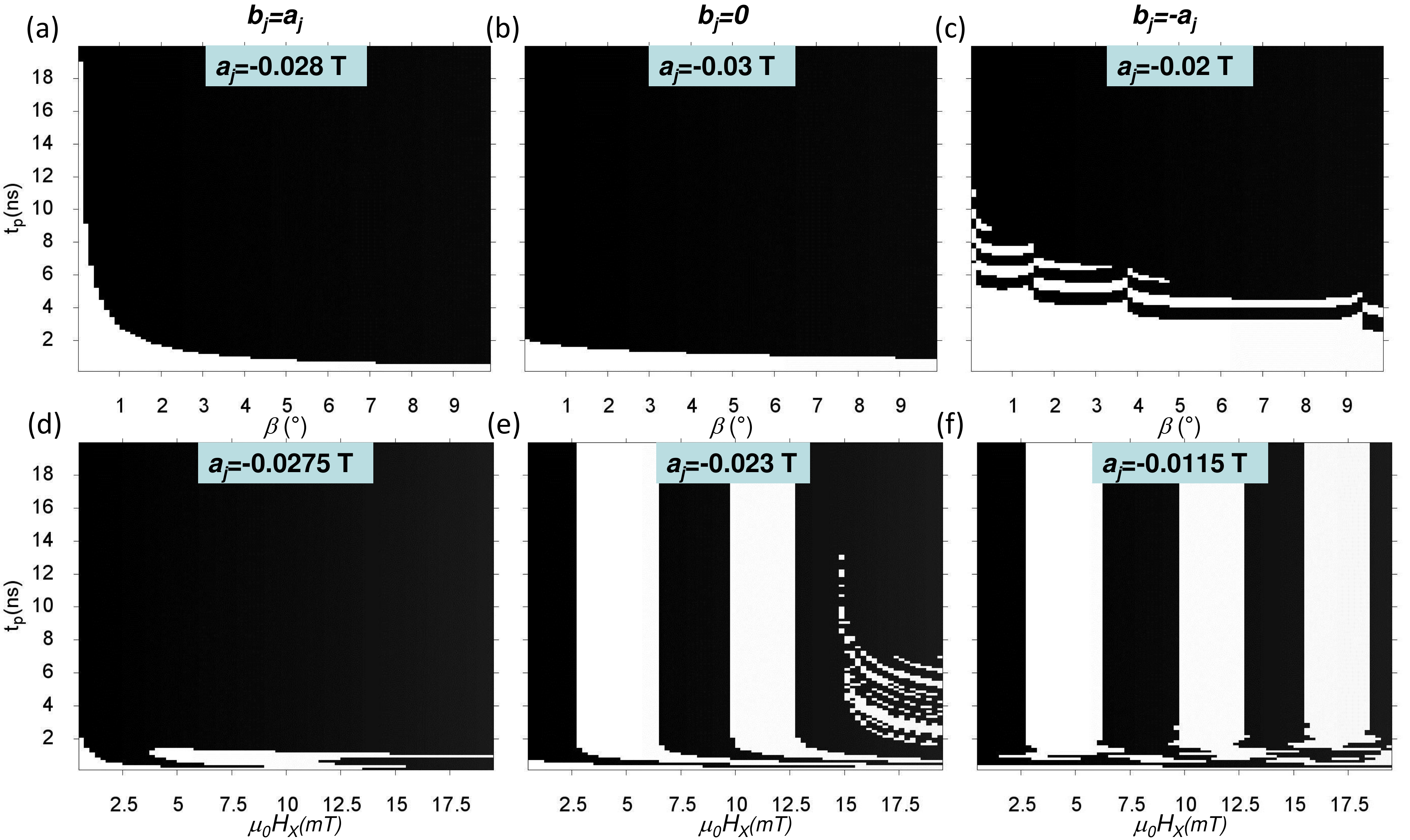}
\caption{SOT switching in the tilt case for different pulse duration $t_p$ as a function of tilt angle $\beta$ for (a) field-torque equal to damping-torque $b_j=a_j$ (b) no field-like torque $b_j=0$ (c) and inverted sign between torques $b_j=-a_j$. In the next row, SOT switching in the bias case for different pulse duration $t_p$ as a function of the bias field $H_x$ for (d) field-torque equal to damping-torque $b_j=a_j$ (e) no field-like torque $b_j=0$ (f) and inverted sign between torques $b_j=-a_j$. Black dot means switched state and white dot not switched.}
\label{Fig:pulse}
\end{figure*}

Another important feature which we need to include in the analysis of hypothetical SOT devices is the ultimate speed of operation. To this end we show here the effect of different current pulse durations $t_p$ on the final magnetization state achieved after a $t= 10 t_p$ relaxation time, see Fig. \ref{Fig:pulse}. The current $a_j$ is fixed at a value where switching is achieved, and close to the critical current $|a_{j,c}|$ and then we vary both the tilt angle $\beta$ or the bias field $H_{X}$ and also the pulse duration $t_p$. After relaxation without applied current for $t= 10 t_p$, we draw a black dot in the graph if the switching takes place or a white one if switching does not occur. 

Fig.\ref{Fig:pulse}(a) to (c) shows switching speed for the tilt case. For both $b_j$=$a_j$ and $b_j$=0 the switching is permanent for a pulse duration $t_p >$ 3 ns. With the exception of the much slower $b_j$=-$a_j$ case, the performance of a tilt based device results are far faster than an AFM/FM stack with bias field. However the fastest speed here is found when $b_j$=0, a condition which is unlikely to be achieved in real samples, due to the fact that the physical phenomena which tend to generate the damping-like torque $b_j$ also tend to produce a field-like one $a_j$ \cite{reviewsot}. 

Fig \ref{Fig:pulse}(d) to (f) shows the bias case. Here the $b_j$=$a_j$ case behaves quite well and manages to be stable. There are some instabilities near zero field, which are expected due to the necessity of perturb the energy landscape with an external field to obtain a field-free switching system. The other two cases $b_j=0$, and $b_j$=$-a_j$ present considerable areas where switching does not occur. Particularly, the $b_j=0$ case is mostly unstable above 15 mT. 

The pulse duration analysis shows that extra care is needed to choose the operating current value. Not only the minimum switching current has to be considered, but also how the system evolves and relaxes with time. As mentioned above, the field-like and damping-like torque have different signs during the precession period, and the specific instant at which they are removed interferes with the trajectory taken by the magnetization $m_z$ to the final state.

\section{Conclusions}

We have analyzed the field-free switching by spin-orbit torques for two different mechanism that yield to a deterministic behavior with HM/FM with tilted anisotropy and AFM/FM bilayers with bias field. While in the tilt case the energy barrier is preserved with increasing anisotropy tilt angles, we have found a significant reduction of the anisotropy energy barrier for the bias case caused by the bias field. For a comparison the Landau-Lifshitz-Gilbert equation was also solved numerically in a macrospin approximation with added damping-like and field-like torques as a function of the variable of interest for each case (tilt angle or bias field). 

Our numerical simulations confirmed the analytical results of a reduction of the energy barrier for the AFM/FM bias field case. On the other hand, in the HM/FM anisotropy tilt case, we have shown the switching to be specially robust with small variations of the tilt angle. Moreover, this method also tolerates a certain flexibility on the chosen HM layer, allowing to employ novel materials with giant spin Hall angles (as the oxygen doped W and Ta) \cite{Derunova2019,liu12,demasius2016,hao2019,zhu2018} and consequently to lower the required current densities. Since most of the other more complex switching methods proposed, and not considered in this paper, also present a coupling with another layer, and hence a potential energy barrier reduction, we thus believe that the tilt case to be a strong contender towards future practical magnetic memory applications. The missing piece is then to find a way to systematically generate a small anisotropy tilt on the FM layer. This may be done, for example, by tweaking the deposition process, introducing an inhomogeneity of the sputtering on the target or causing a mismatch in the crystalline axis of the layers in the interface.

\bibliographystyle{ieeetr}
\bibliography{ref_macrospin4}

\end{document}